\input harvmac
\input graphicx

\def\Title#1#2{\rightline{#1}\ifx\answ\bigans\nopagenumbers\pageno0\vskip1in
\else\pageno1\vskip.8in\fi \centerline{\titlefont #2}\vskip .5in}
%

%%%%%%%%%%%%%%%%%%
%
% Figure macros, SBG 3/03
%
\ifx\includegraphics\UnDeFiNeD\message{(NO graphicx.tex, FIGURES WILL BE IGNORED)}
\def\figin#1{\vskip2in}% blank space instead
\else\message{(FIGURES WILL BE INCLUDED)}\def\figin#1{#1}
\fi
\def\Fig#1{Fig.~\the\figno\xdef#1{Fig.~\the\figno}\global\advance\figno
 by1}
%
%  Ifig   usage:
%
%         \Ifig{\Fig\figlabel}{caption}{figfile}{hsize}
%
% where vsize is the desired vertical size of the figure in truein
%
\def\Ifig#1#2#3#4{
\goodbreak\midinsert
\figin{\centerline{
\includegraphics[width=#4truein]{#3}}}
\narrower\narrower\noindent{\footnotefont
{\bf #1:}  #2\par}
\endinsert
}
%
%defs
%
\font\ticp=cmcsc10

\def\roughly#1{\mathrel{\raise.3ex\hbox{$#1$\kern-.75em\lower1ex\hbox{$\sim$}}}}

%
%refs
%
\lref\hawkrad{
  S.~W.~Hawking,
  ``Particle Creation By Black Holes,''
  Commun.\ Math.\ Phys.\  {\bf 43}, 199 (1975)
  [Erratum-ibid.\  {\bf 46}, 206 (1976)].
  %%CITATION = CMPHA,43,199;%%
}
\lref\SGinfo{S.~B.~Giddings,
  ``Quantum mechanics of black holes,''
  arXiv:hep-th/9412138\semi
  %%CITATION = HEP-TH 9412138;%%
  ``The black hole information paradox,''
  arXiv:hep-th/9508151.
  %%CITATION = HEP-TH 9508151;%%
}
\lref\thooholo{
  G.~'t Hooft,
  ``Dimensional reduction in quantum gravity,''
  arXiv:gr-qc/9310026.
  %%CITATION = GR-QC/9310026;%%
}
\lref\SusskindVU{
  L.~Susskind,
  ``The World As A Hologram,''
  J.\ Math.\ Phys.\  {\bf 36}, 6377 (1995)
  [arXiv:hep-th/9409089].
  %%CITATION = JMAPA,36,6377;%%
}
\lref\BPS{
  T.~Banks, L.~Susskind and M.~E.~Peskin,
  ``Difficulties For The Evolution Of Pure States Into Mixed States,''
  Nucl.\ Phys.\  B {\bf 244}, 125 (1984).
  %%CITATION = NUPHA,B244,125;%%
}
\lref\PreskillTC{
  J.~Preskill,
  ``Do black holes destroy information?,''
  arXiv:hep-th/9209058.
  %%CITATION = HEP-TH/9209058;%%
}
\lref\HaydenCS{
  P.~Hayden and J.~Preskill,
  ``Black holes as mirrors: quantum information in random subsystems,''
  JHEP {\bf 0709}, 120 (2007)
  [arXiv:0708.4025 [hep-th]].
  %%CITATION = JHEPA,0709,120;%%
}
\lref\PageUP{
  D.~N.~Page,
  ``Black hole information,''
  arXiv:hep-th/9305040.
  %%CITATION = HEP-TH/9305040;%%
}
\lref\wabhip{
  S.~B.~Giddings,
  ``Why Aren't Black Holes Infinitely Produced?,''
  Phys.\ Rev.\  D {\bf 51}, 6860 (1995)
  [arXiv:hep-th/9412159].
  %%CITATION = PHRVA,D51,6860;%%
}
\lref\susstrouble{
  L.~Susskind,
  ``Trouble For Remnants,''
  arXiv:hep-th/9501106.
  %%CITATION = HEP-TH/9501106;%%
}
\lref\pairs{
  L.~Parker,
  ``Probability Distribution Of Particles Created By A Black Hole,''
  Phys.\ Rev.\  D {\bf 12}, 1519 (1975).
  %%CITATION = PHRVA,D12,1519;%%
}
\lref\GGP{
  M.~Gary, S.~B.~Giddings and J.~Penedones,
  ``Local bulk S-matrix elements and CFT singularities,''
  arXiv:0903.4437 [hep-th].
  %%CITATION = ARXIV:0903.4437;%%
}
\lref\GMH{
  S.~B.~Giddings, D.~Marolf and J.~B.~Hartle,
  ``Observables in effective gravity,''
  Phys.\ Rev.\  D {\bf 74}, 064018 (2006)
  [arXiv:hep-th/0512200].
  %%CITATION = PHRVA,D74,064018;%%
}
\lref\GaGi{
  M.~Gary and S.~B.~Giddings,
  ``Relational observables in 2d quantum gravity,''
  arXiv:hep-th/0612191, Phys.\ Rev.\  D {\bf 75} 104007 (2007).
  %%CITATION = HEP-TH/0612191;%%
}
\lref\GiSr{
  S.~B.~Giddings and M.~Srednicki,
  ``High-energy gravitational scattering and black hole resonances,''
  Phys.\ Rev.\  D {\bf 77}, 085025 (2008)
  [arXiv:0711.5012 [hep-th]].
  %%CITATION = PHRVA,D77,085025;%%
}
\lref\GiPo{
 S.~B.~Giddings and R.~A.~Porto,
  ``The gravitational S-matrix,''
  Phys.\ Rev.\  D {\bf 81}, 025002 (2010)
  [arXiv:0908.0004 [hep-th]].
  %%CITATION = ARXIV:0908.0004;%%
}
\lref\BHMR{
  S.~B.~Giddings,
  ``Black holes and massive remnants,''
  Phys.\ Rev.\  D {\bf 46}, 1347 (1992)
  [arXiv:hep-th/9203059].
  %%CITATION = PHRVA,D46,1347;%%
}
\lref\Astroprob{
  A.~Strominger,
  ``Five Problems in Quantum Gravity,''
  Nucl.\ Phys.\ Proc.\ Suppl.\  {\bf 192-193}, 119 (2009)
  [arXiv:0906.1313 [hep-th]].
  %%CITATION = NUPHZ,192-193,119;%%
}
\lref\STU{
  L.~Susskind, L.~Thorlacius and J.~Uglum,
  ``The Stretched Horizon And Black Hole Complementarity,''
  Phys.\ Rev.\  D {\bf 48}, 3743 (1993)
  [arXiv:hep-th/9306069].
  %%CITATION = PHRVA,D48,3743;%%
}
\lref\LPSTU{
  D.~A.~Lowe, J.~Polchinski, L.~Susskind, L.~Thorlacius and J.~Uglum,
  ``Black hole complementarity versus locality,''
  Phys.\ Rev.\  D {\bf 52}, 6997 (1995)
  [arXiv:hep-th/9506138].
  %%CITATION = PHRVA,D52,6997;%%
}
\lref\GiNe{
  S.~B.~Giddings and W.~M.~Nelson,
  ``Quantum emission from two-dimensional black holes,''
  Phys.\ Rev.\  D {\bf 46}, 2486 (1992)
  [arXiv:hep-th/9204072].
  %%CITATION = PHRVA,D46,2486;%%
}
\lref\Page{
  D.~N.~Page,
  ``Information in black hole radiation,''
  Phys.\ Rev.\ Lett.\  {\bf 71}, 3743 (1993)
  [arXiv:hep-th/9306083].
  %%CITATION = PRLTA,71,3743;%%
}
\lref\QBHB{
  S.~B.~Giddings,
  ``Quantization in black hole backgrounds,''
  Phys.\ Rev.\  D {\bf 76}, 064027 (2007)
  [arXiv:hep-th/0703116].
  %%CITATION = PHRVA,D76,064027;%%
}
\lref\Hawkunc{
  S.~W.~Hawking,
  ``Breakdown Of Predictability In Gravitational Collapse,''
  Phys.\ Rev.\  D {\bf 14}, 2460 (1976).
  %%CITATION = PHRVA,D14,2460;%%
}
\lref\thooconf{
  G.~'t~Hooft,
  ``Quantum gravity without space-time singularities or horizons,''
  [arXiv: 0909.3426 [gr-qc]].
  %%CITATION = ARXIV:0909.3426;%%
}
\lref\GiddingsNC{
  S.~B.~Giddings and M.~S.~Sloth,
  ``Semiclassical relations and IR effects in de Sitter and slow-roll
  space-times,''
  arXiv:1005.1056 [hep-th].
  %%CITATION = ARXIV:1005.1056;%%
}
\lref\tHooftRE{
  G.~'t Hooft,
  ``On The Quantum Structure Of A Black Hole,''
  Nucl.\ Phys.\  B {\bf 256}, 727 (1985).
  %%CITATION = NUPHA,B256,727;%%
}
\lref\StWi{
  R.~F.~Streater and A.~S.~Wightman,
  ``PCT, spin and statistics, and all that,''
%\href{http://www.slac.stanford.edu/spires/find/hep/www?irn=2423855}{SPIRES entry}
{\it  Redwood City, USA: Addison-Wesley (1989) 207 p. (Advanced book classics).}
}
\lref\Sussscramb{
  Y.~Sekino and L.~Susskind,
  ``Fast Scramblers,''
  JHEP {\bf 0810}, 065 (2008)
  [arXiv: 0808.2096 [hep-th]].
  %%CITATION = JHEPA,0810,065;%%
}
\lref\GaGiAdS{
  M.~Gary and S.~B.~Giddings,
  ``The flat space S-matrix from the AdS/CFT correspondence?,''
  Phys.\ Rev.\  D {\bf 80}, 046008 (2009)
  [arXiv:0904.3544 [hep-th]].
  %%CITATION = PHRVA,D80,046008;%%
}
\lref\HPPS{
  I.~Heemskerk, J.~Penedones, J.~Polchinski and J.~Sully,
  ``Holography from Conformal Field Theory,''
  arXiv:0907.0151 [hep-th].
  %%CITATION = ARXIV:0907.0151;%%
}
\lref\LQGST{
  S.~B.~Giddings,
  ``Locality in quantum gravity and string theory,''
  Phys.\ Rev.\  D {\bf 74}, 106006 (2006)
  [arXiv:hep-th/0604072].
  %%CITATION = PHRVA,D74,106006;%%
}
\lref\GGM{
  S.~B.~Giddings, D.~J.~Gross and A.~Maharana,
  ``Gravitational effects in ultrahigh-energy string scattering,''
  Phys.\ Rev.\  D {\bf 77}, 046001 (2008)
  [arXiv:0705.1816 [hep-th]].
  %%CITATION = PHRVA,D77,046001;%%
}
\lref\GiddingsSJ{
  S.~B.~Giddings,
  ``Black hole information, unitarity, and nonlocality,''
  Phys.\ Rev.\  D {\bf 74}, 106005 (2006)
  [arXiv:hep-th/0605196].
  %%CITATION = PHRVA,D74,106005;%%
}
\lref\GiddingsBE{
  S.~B.~Giddings,
  ``(Non)perturbative gravity, nonlocality, and nice slices,''
  Phys.\ Rev.\  D {\bf 74}, 106009 (2006)
  [arXiv:hep-th/0606146].
  %%CITATION = PHRVA,D74,106009;%%
}
\lref\GiddingsPJ{
  S.~B.~Giddings,
  ``Black holes, information, and locality,''
  Mod.\ Phys.\ Lett.\  A {\bf 22}, 2949 (2007)
  [arXiv:0705.2197 [hep-th]].
  %%CITATION = MPLAE,A22,2949;%%
}
\lref\SusskindMU{
  L.~Susskind and L.~Thorlacius,
  ``Gedanken experiments involving black holes,''
  Phys.\ Rev.\  D {\bf 49}, 966 (1994)
  [arXiv:hep-th/9308100].
  %%CITATION = PHRVA,D49,966;%%
}
\lref\Mathur{
  S.~D.~Mathur,
  ``The information paradox: A pedagogical introduction,''
  Class.\ Quant.\ Grav.\  {\bf 26}, 224001 (2009)
  [arXiv:0909.1038 [hep-th]].
  %%CITATION = CQGRD,26,224001;%%
}
\Title{
\vbox{%\hbox{Draft -- do not distribute}
\hbox{CERN-PH-TH/2009-221}}
\vbox{\baselineskip12pt
}}
{\vbox{\centerline{Nonlocality vs. complementarity: a conservative}\centerline{ approach to the information problem}
}}
\centerline{{\ticp Steven B. Giddings\footnote{$^\ast$}{Email address: giddings@physics.ucsb.edu}  } }
\centerline{\sl Department of Physics}
\centerline{\sl University of California}
\centerline{\sl Santa Barbara, CA 93106}
\vskip.05in
\centerline {and}
\vskip.05in
\centerline{\sl PH-TH, CERN}
\centerline{\sl 1211 Geneve 23, Switzerland}
\vskip.10in
\centerline{\bf Abstract}
A proposal for resolution of the information paradox is that ``nice slice" states, which have been viewed as providing a sharp argument for information loss, do not in fact do so as they do not give a fully accurate description of the quantum state of a black hole.   
This however leaves an information {\it problem}, which is to provide a consistent description of how information escapes when a black hole evaporates.  While a rather extreme form of nonlocality has been advocated in the form of complementarity, this paper argues that is not necessary, and more modest nonlocality could solve the information problem. One possible distinguishing characteristic of scenarios is the information retention time.  The question of whether such nonlocality implies acausality, and particularly inconsistency, is briefly addressed. The need for such nonlocality, and its apparent tension with our empirical observations of local quantum field theory, may be a critical missing piece in understanding the principles of quantum gravity.

\vskip.3in
%\draftmode
\Date{}

The discovery of black hole evaporation\refs{\hawkrad} catalyzed emergence of the information paradox.\foot{For reviews, see \refs{\PageUP,\SGinfo}.}  Perspectives differ on the status of this paradox, ranging from beliefs it arises from a simple mistake, to viewing it as a problem as fundamentally important as that of the classical  instability of the hydrogen atom\refs{\GiddingsSJ\GiddingsBE-\GiddingsPJ}.

The trouble arises when considering the fate of information that falls into a black hole.  Hawking's arguments\hawkrad\ and subsequent improvements tell us that, by locality, it cannot escape during evaporation.  Difficulties in modifying quantum mechanics, and particularly resultant drastic violation of energy conservation\refs{\BPS}, tell us it cannot be destroyed.  And, instabilities to infinite pair production\refs{\tHooftRE\PreskillTC-\wabhip} and difficulties with virtual effects\refs{\susstrouble} tell us it cannot be left behind in a remnant.  This contradiction between principles is the essence of the paradox.

If a fundamental principle requires revision, plausibly the weakest candidate in quantum gravity is locality.  As a cornerstone of quantum field theory, locality is encoded in commutativity of local observables outside the light cone, but there are no such gauge invariant observables in quantum gravity, and at best they only emerge in an approximation\refs{\GMH,\GaGi}.  Locality can also be probed through high-energy behavior of the S-matrix, but it has  been argued that in gravity this behavior is unusual from the local field theory perspective, and plausibly not local\refs{\GiSr,\GiPo}.  

One early proposal for a nonlocal resolution of the problem\refs{\BHMR}\foot{See also \refs{\Astroprob} for a similar proposal.  The earlier proposal lay dormant due to criticisms regarding causality; see further discussion later in this paper.} suggested that bounds on information content in a region resulted in nonlocal propagation of information at spacelike separations with respect to the semiclassical geometry of a black hole, allowing its escape, and possibly yielding a massive remnant. 

Another more widely discussed proposal is that of black hole {\it complementarity}.  In the form advocated in \STU, this states that observations of observers who stay away from the black hole, and of those who fall in, are complementary in an analogous fashion to complementarity of observations of $x$ and $p$ in quantum mechanics, and thus should not simultaneously enter a physical description.  This proposal is closely associated with that of {\it holography}\refs{\thooholo,\SusskindVU}, which in its strong form states that for a given region of space, like that inside a black hole, there is a completely equivalent description in terms of a lower-dimensional field theory on its boundary.  Complementarity represents a rather extreme departure from locality, in forbidding simultaneous description of information separated by large spacelike intervals.  

The purpose of this paper will be to more closely examine the motivations for  different kinds of nonlocality, and ultimately to argue that a less radical version could restore unitarity in black hole evaporation.  Indeed, one important characteristic of proposals for how information escapes is the information retention time, $T_{ret}$.  For a black hole with initial horizon radius $R$, this has in some discussions (see {\it e.g.} \HaydenCS) been assumed to be of magnitude 
\eqn\compret{T_{ret}\sim R\log R\ ,}
partly motivated by complementarity arguments.  In the leading semiclassical approximation the retention time is longer than the lifetime of the black hole (the information never escapes\Hawkunc).  One might question whether a comparatively short retention time \compret\ is logically necessary, or whether the actual retention time could be significantly longer.  Moreover, it is important to address the question of whether some nonlocal information escape at any time scale leads to basic inconsistencies, due to {\it e.g.} acuasality\thooconf.  This paper examines both of these issues more closely, ultimately arguing for the possible logical consistency of a longer retention time, without  acausality leading to inconsistency.

To understand motivations for complementarity, let us first review Hawking's original argument, in its most advanced version: the nice slice argument.  Consider a black hole that has formed from a collapsing body whose details won't concern us.  Suppose that we inject another quantum into the black hole.  We think of it as a qubit; its quantum information could be carefully tracked by describing its correlations with other quanta.\foot{Similar statements can be made about the Hawking radiation itself, which can be described in terms of correlated pairs of particles\refs{\pairs,\GiNe}, one member of which stays within the black hole.}  With unitary evolution, the information of this bit will eventually be radiated in Hawking radiation.  This information may be very scrambled, and come out in quite subtle correlations in multi-particle states.  But, let $T_{ret}$ be the time at which an ${\cal O}(1)$ fraction of this information has been reemitted.  Now, we may draw a spatial slice, asymptoting to a slice of constant Schwarzschild time $t>T_{ret}$, and which dives into the horizon, but then asymptotes to a slice of constant Schwarzschild $r=r_c$, with $r_c$ inside the horizon radius $R$, but $r_c\gg l_{Pl}$.  Thus, this {\it nice slice} is spacelike, and moreover crosses the infalling qubit in a region far away from the singularity, as shown in Fig.~1.  

\Ifig{\Fig\figEFns}{A Penrose diagram for an evaporating black hole, showing an infalling qubit, its information that might escape on a timescale $T_{ret}$, and a nice slice as described in the text.}{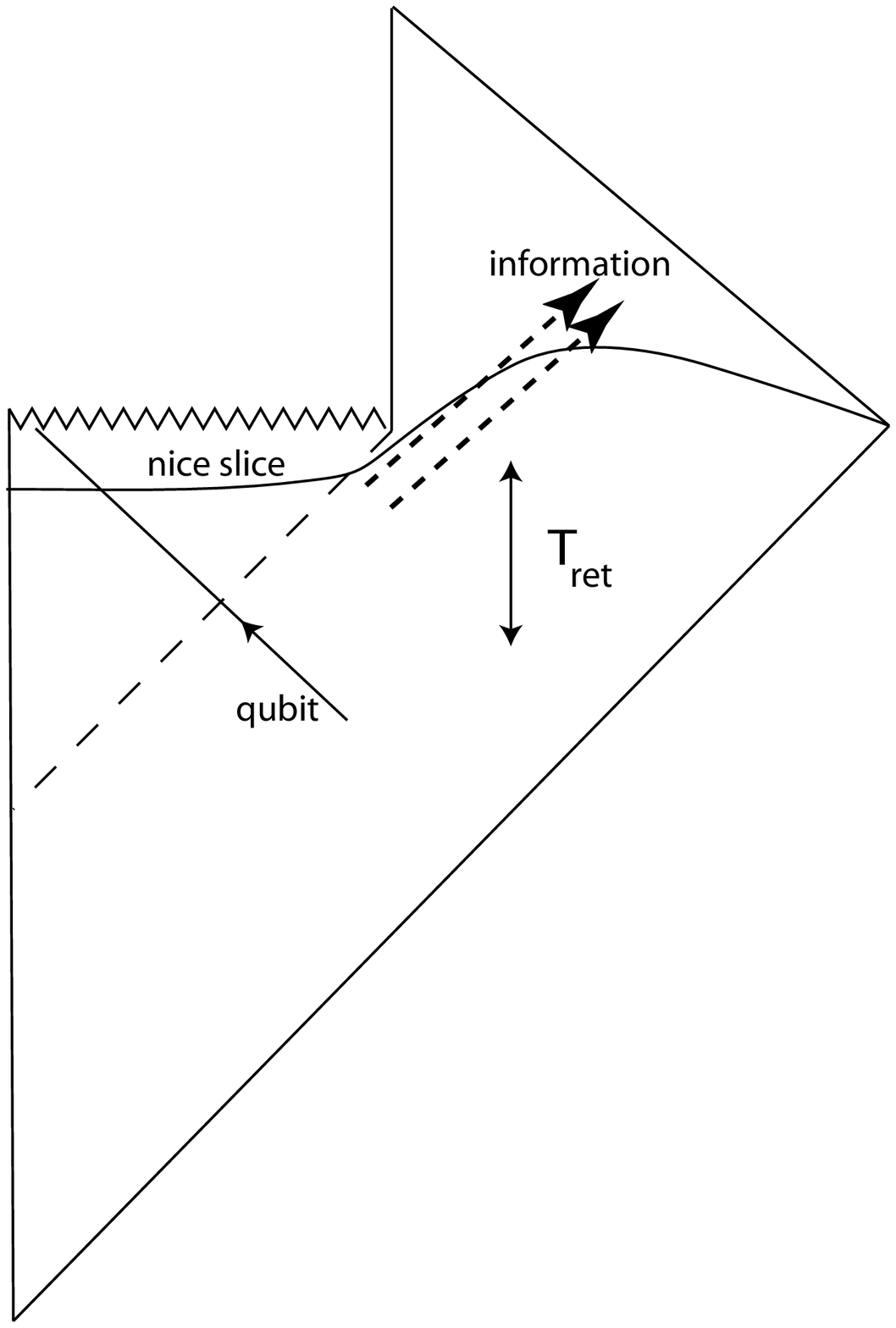}{3}

Since curvatures are small, nothing interferes with the qubit preserving its information  to where it intersects the slice.  But, by locality, this information is independent of the information in the spatially-separated Hawking radiation, and cannot be transmitted there.  Moreover, the no-cloning principle of quantum mechanics tells us it cannot be duplicated there.   Thus, escape of information from the black hole is actually prohibited by locality.  This argument apparently only breaks down if $T_{ret}$ is late enough that the black hole becomes planckian in size, but if so, it would take a much {\it longer} time for the information to escape, yielding a remnant scenario.  In fact, Page\refs{\Page} has argued that in order for the information to escape a black hole in Hawking radiation,  $T_{ret}$ must less than of order the half-life of the black hole.  This gives, parametrically,
\eqn\tub{T_{ret}\roughly< RS,}
where $S$ denotes the Bekenstein-Hawking entropy.

The complementarity proposal states that physics forbids simultaneous description of the information inside and outside the black hole.  A particular example
would be given by a quantum state on the nice slice.  If one takes the outside observer's viewpoint and thus forbids any simultaneous inside description, one evades contradiction from information escape.  This naturally fits with holography, in that the description of the entire interior of the black hole might then be replaced by a quantum description at its horizon surface, from which information could be emitted.  Indeed, it has been advocated that outside observers see an infalling observer ``burn up" (thermalize), whereas the infalling observer makes complementary observations in which nothing unusual happens at the horizon.

One might imagine finding a contradiction if an observer who stays outside the black hole past $T_{ret}$ and captures the information then falls into the black hole, and might then be able to compare the captured information with that in the infalling qubit, contradicting the prohibition on quantum cloning.  To answer this, \SusskindMU\ shows that if the retention time satisfies 
\eqn\retlb{T_{ret}\roughly> R \log R,}
it is not possible for a signal to be sent from the qubit to the later-infalling observer without using ultraplanckian energies.

What is not clear is that physics demands such a radical departure from local quantum field theory, which allows a perfectly good description of both the exterior and the interior of the black hole, outside of the strong-curvature regime.  Of course, taken to an extreme, such a description also yields the nice slice argument -- but an important point, which we now turn to, is that this {\it is} an extreme.

\Ifig{\Fig\figEF}{An Eddington-Finkelstein diagram of the black hole.  The vertical dashed line is the horizon, at Schwarzschild coordinate $r=R$.  Also shown are an infalling qubit, its outgoing information possibly emitted after $T_{ret}\gg R$, and three members of a family of nice slices.}{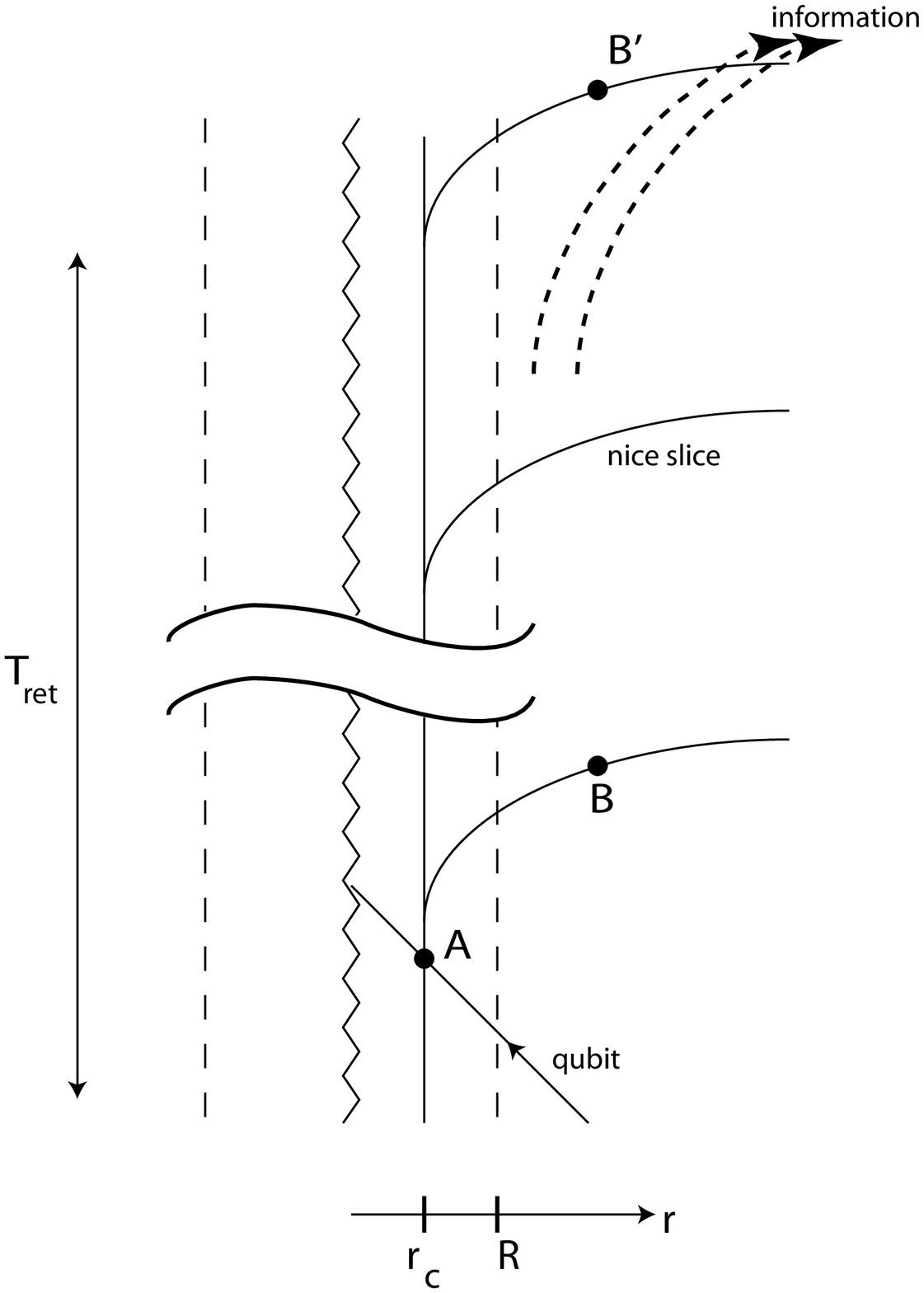}{4}

To see this, first note that, following \refs{\LPSTU,\QBHB,\Mathur}, a family of nice slices may be constructed by performing Schwarzschild time translations of the slice already constructed.  Since these are symmetries of the metric, the intrinsic geometry of the slice doesn't change.\foot{When one takes into account the slow decrease of the mass, there are corrections to this statement, to which we will return.}  However, there are two notable aspects of the translation.  First, the distance from a given spacetime point at $r=r_c$ on a slice (say, point $A$ in \figEF), to the point where the slice intersects a given external radius, say $r=2R$ (point $B$ in \figEF), increases by an amount 
\eqn\lincrease{\Delta l= \sqrt{R/r_c-1}\Delta t}
under translation by time $\Delta t$ -- the slices lengthen, by adding segments along $r=r_c$.  Moreover, in the vicinity of the horizon, Schwarzschild time translations act as boosts, with boost parameter $\Delta \theta =  \Delta t/2R$.   More general constructions of such nice slices should have the same basic features, namely accumulating at a radius outside the strong curvature region, and lengthening with time evolution.

To illustrate how extreme such a construction is, note  that the spatial distance along the slice between the impact of the infalling qubit and the outgoing information (near $B^{\prime}$ in \figEF) is of size
\eqn\lret{L_{ret}\approx \sqrt{R/r_c-1} T_{ret}.}
For example, using the Page bound \tub\ gives a spatial separation $L_{ret}\sim R S$.  For a solar mass black hole, with $R\sim km$, this gives  $L_{ret}\sim 10^{80} m$!  However, along other paths between $A$ and $B'$, the spatial separation is as small as $\sim R\sim km$.

A sharp calculation of the missing information in Hawking radiation, updating \refs{\Hawkunc}, can be attempted based on construction of the quantum state on the nice slice, and can be sketched as follows.  First, one can compute the state $|\psi\rangle$ on the nice slice labelled by time $t$, given specification of an initial state for the collapse problem; for a concrete two-dimensional example, see {\it e.g.} \refs{\GiNe}.  
Next, by locality, one can write this state in terms of states in a product of Hilbert spaces interior and exterior to the black hole.  The exterior state of the black hole is thus described by a density matrix, $\rho = {\rm Tr}_{in} (|\psi\rangle\langle\psi|).$  The entropy of this density matrix, parametrizing the missing information, is $S_{miss}=-{\rm Tr} (\rho \log \rho)$.  If one follows this construction to a time where most of the mass of the black hole has been radiated, one finds $S_{miss}$ is of size the Bekenstein-Hawking value $S$.  While the two-dimensional example nicely illustrates this result, one can outline arguments for its generalization to the higher-dimensional case.  One approach to this is to note that Hawking radiation is dominated by the lowest orbital angular momentum states, for example s-wave for scalars; evolution of these states is effectively two-dimensional.  Alternately, we expect that to each Hawking quantum there is associated one missing bit of information, resulting from a correlated partner quantum falling into the black hole.  Since the number of such quanta emitted is $\sim S$, this is the approximate size of the missing information.\foot{Both of these statements are expected to be confirmed by a detailed analysis of the higher-dimensional case, generalizing \refs{\GiNe}.} 

We now reach one of the central points of the discussion.  The extreme nature of the nice slice construction suggests that a semiclassical nice slice state like we have just described is not a fully accurate representation of the quantum state of the black hole.  But, without a sharp calculation of the missing information, which we have based on this state, there is no paradox.  

There are at least two reasons\refs{\QBHB} to question whether a nice slice state gives a sufficiently accurate description of the dynamics, and plausibly more.

First, note that it is very hard to give a careful definition of what one means by such a state, and even harder to give an operational definition of how such a state could be measured.
Specification of such a state is a very gauge-dependent construct, whose gauge-independent version is very subtle to give.  The latter relies on having a way to describe local quantum information in a diffeomorphism-invariant fashion, and runs directly into the well-known lack of local gauge-invariant observables in gravity.  There are  proposals on how some such approximately local observables can be formulated, in \refs{\GMH,\GaGi}.  These follow a relational approach, in which one must include other fields which provide reference backgrounds and/or detectors, with respect to which the state can be described in a diffeomorphism-invariant fashion.  From a less-technical perspective, one could simply ask the related question of how such a state could be measured. 

The ``missing information" in the state can be thought of as largely due to correlated partners of outgoing Hawking particles\refs{\pairs,\GiNe,\QBHB}.  These have a characteristic wavelength $\sim R$, and roughly one of them impacts the nice slice in each time interval $\sim R$.  Thus, to measure this structure in the nice slice state, one might try to construct a constellation of detectors with separations $\sim R$, which free-fall into the black hole and reach the appropriate segment of the nice slice we would like to measure.  However, the minimum energy of such a detector is clearly $1/R$.  The number of detectors needed (assuming \tub) is $N\sim T_{ret}/R\sim S$, and the total energy of the constellation is $E\roughly>M$. Thus, the nice slice state cannot be carefully measured, without producing a large perturbation on the state.  Moveover, even if such a collection of detectors measured the state, there is no way for their observations to be compared at a central location -- radio signals that they send forward in time enter the high-curvature region of the black hole without meeting.\foot{An argument of this second form was used as motivation for complementarity\refs{\STU,\SusskindMU}.}  

This suggests that a nice slice state has no sharp meaning, as a complete description of the quantum black hole, based on a minor modification of a dictum due to Bohr and Wheeler:  ``no phenomenon is a real phenomenon unless it can in principle be observed."  For those suspicious of such positivist arguments, note that this objection to accuracy of nice-slice states is actually a much more mild application of such reasoning than that used in justifying complementarity\refs{\STU,\SusskindMU}!

A second problem with nice-slice states arises when one ignores the preceding  objection and  simply tries to carefully justify the leading semiclassical approximation through an expansion in a small parameter.  In particular, corrections to the leading semiclassical approximation include the backreaction on the state due to perturbations, say from  fluctuations in the Hawking radiation, or additional quanta falling in\QBHB.  Hawking quanta have a typical energy $E\sim 1/R$, and ordinarily such quanta would produce absolutely negligible effect.  However,  as in the example explained in \QBHB, such perturbations can be leveraged by the very long time scale $T_{ret}$ into an ${\cal O}(1)$ effect on the nice slice state.\foot{As outlined in \refs{\QBHB,\GiddingsNC}, there are related effects in accelerating cosmologies on time scales $\sim R_{dS} S_{dS}$.}  Thus, at the level of detail needed to calculate $S_{miss}$, there is a source of instability and uncertainty in the nice slice state.  More careful calculations are needed, but it is not clear this can be overcome.

As noted, there may be other issues found with more careful examination of nice slice states; in any case the preceding discussion suggests the first part of the proposed answer\QBHB\ to the information question:

\item{1.} {\it The nice slice construction does not describe the state of the black hole precisely enough to sharply calculate the missing information.  Without such a sharp calculation, there is no paradox.}

Such a  resolution of the paradox leaves an information {\it problem} -- namely, what quantum dynamics gives a means of calculating the unitary evolution, with return of the information, {\it e.g.} in modified Hawking radiation?  A plausible viewpoint is that the need for a quantum theory of gravity to be unitary in these contexts should be an essential guide towards the mechanisms, principles, and mathematical formalism of that theory.   Without yet understanding these principles, one can ask whether there is a reasonable description of a scenario, with minimal departure from the semiclassical picture, and that  is less radical than complementarity.\foot{'t Hooft has recently proposed another variant of complementarity\refs{\thooconf}; however, this also apparently represents a large departure from the semiclassical local description  of an evaporating black hole.}  Referring to the analogy of quantum mechanics, such a semiclassical picture may ultimately be at least as fundamentally inaccurate as the description of electron dynamics within the  Bohr radius by classical mechanics.  But a minimum criterion for reasonableness is, at least, logical consistency.

\Ifig{\Fig\natslice}{A different slicing of the black hole geometry, by ``natural" slices.  Also shown are the coordinate axes of Eddington-Finkelstein coordinates $v$ and $r$. }{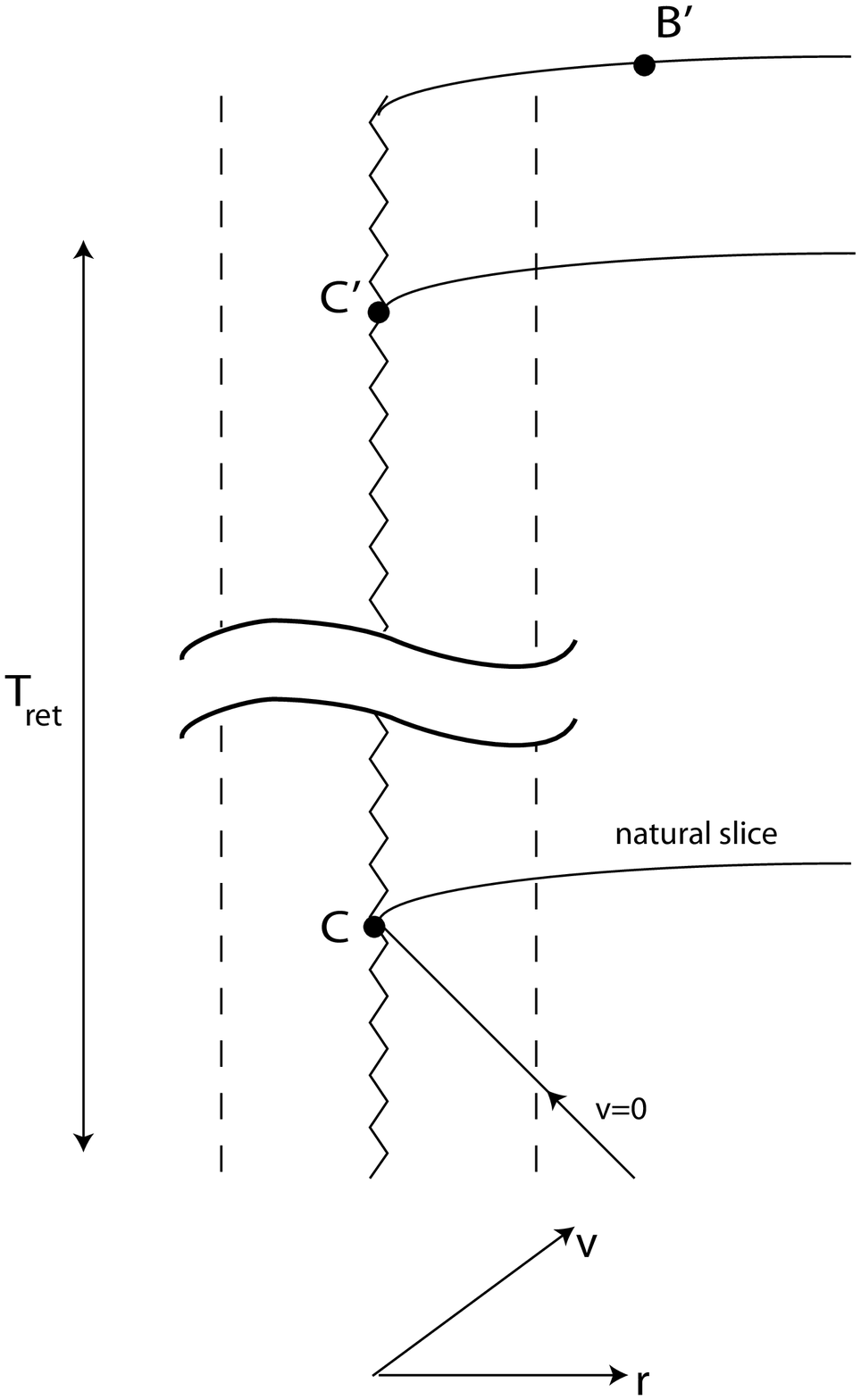}{4}

A second central point of the discussion is that the information problem could be resolved by some modest nonlocality of quantum gravity, with respect to the semiclassical picture.  To approach this point, first let us examine what might be a more reasonable description of evolution inside the black hole, given the preceding objections to nice slices.  As an example, one might start with our constellation of free-falling satellites, distributed in $r$, and with initially synchronized clocks, which  define spatial slices and quantum states at a given common time with respect to their clocks.  The innermost satellites fall into the black hole, and ultimately into the strong curvature region.  Thus corresponding slices, pictured in \natslice, terminate in this region.  These slices do not appear to suffer from the same objections to the artificially-constructed nice slices, and might be referred to as {\it natural slices}.\foot{Another common slicing is one with spatial slices remaining outside the horizon, like constant Schwarzschild time slices.  Such slices are neither nice nor natural; as is well known observers traveling orthogonally to these slices, along lines of constant $r$, must be extremely accelerated near the horizon, and infalling matter is highly boosted with respect to such observers.  Even so, it is not clear that a physical description of black hole dynamics in such a gauge realizes the picture where infalling matter is rapidly thermalized at the horizon, in any sharp observational sense.}

In such a slicing, the infalling qubit enters the classically singular strong curvature region at finite time.  A first point to note is that propagation from where it does so, at point $C$ in \natslice, to point $C'$ in the figure, is something that may commonly be regarded as not necessarily troublesome, since it involves the unknown dynamics resolving the singularity; one might simply regard the missing information as ``at" the singularity.
However, note that, with respect to the semiclassical geometry, this would be propagation across large spacelike separations, with distance $\propto T_{ret}$, and hence very nonlocal.  But, such spacelike propagation is plausibly not objectionable if not through large flat regions.
If such ``nonlocality" is permitted in strong curvature regions, it could be that related effects permit propagation from $C$ to $B'$, {\it e.g.} by first propagating to $C'$ and then over the remaining distance $\sim R$.  Indeed, while field theory is not necessarily being advocated as the proper framework for a full description, it should be noted that within field theory, nonlocalities at one scale imply nonlocalities at other scales\refs{\StWi}. 

Moreover, while such propagation that appears nonlocal with respect to the semiclassical description appears unusual, it is not clear that it would contradict any other fundamental principles.  Aside from the propagation ``along the singularity," what is needed is propagation over a distance $\sim R$, on the potentially long timescale $T_{ret}$.  

Thus, one is lead to the second part of the proposal, outlining how the information {\it problem} is solved: 

\item{2.}  {\it Gravitational scattering is unitarized by nonperturbative mechanisms which appear to describe modest nonlocality with respect to a semiclassical picture, and in particular over distances $\sim R$ on timescales $\sim T_{ret}$ in a black hole.}

%The non-perturbative dynamics that unitarizes gravitational scattering
%has  modest nonlocality with respect to a semiclassical picture

There are several further comments to make on this proposal.  First, one does not obviously need to commit to a particular value  of $T_{ret}$.  However, a value at the lower bound \retlb, motivated by the complementarity proposal, and corresponding to ``fast scrambling" of information\refs{\Sussscramb}, is most radically different from the semiclassical picture, which predicts $T_{ret}=\infty$.  If information return is due to small corrections, the smallest  possible correspond to \tub;   slow leaking is not as extreme as fast scrambling.

Second, it seems plausible that the information could be returned through nonlocal interactions that make subtle modifications to the outgoing Hawking radiation.\foot{Such interactions would have to evade constraints including those of \Mathur.}  However, some may instead regard it as likely that information return is accompanied by a more obvious disruption of the state near the horizon.  Refs.~\refs{\BHMR,\Astroprob}  suggested this as a possibility, and the fuzzball idea\refs{\Mathur} appears to fit within the outlines of such a proposal, if it were to somehow give a picture in which an initial vacuum  black hole transitions to a massive fuzzball.  A sharp test distinguishing scenarios is to determine whether an infalling observer experiences anything unusual or violent when crossing the horizon. No disruption is evidently the most conservative scenario, unless there prove to be contrary arguments.

Third, while it has been argued (see {\it e.g.} \refs{\LPSTU}) that string extendedness is the source of nonlocality necessary to resolve the information problem, closer examination\refs{\LQGST,\GGM} suggests this is not the case, and that such nonlocality plausibly instead originates in nonperturbative gravitational dynamics.

Finally, an immediate concern with nonlocality is that it potentially implies acausality.  Nonlocal scenarios are objected to on these grounds in
 \refs{\thooconf}; this concern had also been recognized in \BHMR.
This seems a very general issue, in a flat background -- a Lorentz transformation can convert spacelike propagation of information into propagation backwards in time.  While this would still be outside the lightcone, iterating such propagation can permit propagation into the past light cone, and such acausalities can apparently be used to formulate paradoxes implying inconsistency.

While there is not a complete answer to this without better knowledge of the dynamics, note that the picture of nonlocal propagation proposed in the black hole geometry does {\it not} obviously produce such paradoxes.  Specifically, the symmetry group --  time translations, and rotations for a Schwarzschild black hole -- is much more restricted.  The nonlocal propagation we have described is not converted into acausal propogation by these transformations.  

Such propagation can be perfectly causal with respect to the outside Schwarzschild time -- the missing information comes out with a time {\it delay}, not {\it advance}.\foot{In a related development, it has been argued\refs{\GiPo} that assuming gravity has an S-matrix, it should correspondingly lack unphysical acausal behavior.}  For example, if the information always propagates forward in the advanced time $v$ of the diagram \natslice, and only reemerges at $r\sim R$ at long times, this will be causal and time-delayed with respect to observers in the asymptotically flat region; an illustrative example is propagation along the curve $r=Rv/T_{ret}$.  It is true that such propagation is backward in time, with respect to some inside observers, and can be spacelike in the immediate vicinity of the horizon.  However, if the information is delocalized, and/or is not something that these observers can control and use for signaling, there is no obvious way to produce a paradox.

Of course, a very important question is whether a theory can be constructed,\foot{While it may be that string theory ultimately has such features, perturbative string theory appears to fail in the regime of question.  It is moreover an outstanding question whether non-perturbative formulations of string theory via dualities such as AdS/CFT give a complete answer to the question of gravitational scattering\refs{\GGP\GaGiAdS-\HPPS}, and if they do, it is important to understand the nonlocal mechanisms by which they would address the information problem.} which lacks the requisite locality, but does not produce other nonlocality that leads to inconsistency, for example by permitting spacelike signaling in a Minkoswki background.  It may be that our semiclassical mental picture of a black hole is, as with the classical mechanics of the hydrogen atom, not accurate for detailed questions about the ``location" of all  information, but that the needed modifications do not produce any such inconsistencies.  However, this is clearly a critical tension.  If, for reasons that should be apparent, we refer to the underlying theory as ``non-local mechanics," an essential question is how such a mechanics and accompanying mathematical description can reduce to local quantum field theory in  known low energy circumstances, if it does not have the same intrinsic notion of locality. This quite plausibly results from what is presently an incorrect picture of spacetime.  Such a theory -- which might be regarded as a relatively {\it conservative} answer to the information question -- must be ``nearly local,"  in appropriate circumstances, and certainly must be consistent.  These critical questions of how to obtain ``locality without locality," and consistency, may serve as such  strong constraints that they furnish a useful guide toward formulating the theory.  

\bigskip\bigskip\centerline{{\bf Acknowledgments}}\nobreak
I  greatly appreciate the stimulating atmosphere and hospitality of the theory group at CERN, where this paper was written, and thank R. Porto for comments.
This work  was supported in part by the Department of Energy under Contract DE-FG02-91ER40618.

\listrefs
\end